\documentclass[10pt, prl, aps, twocolumn,showpacs,nofootinbib,longbibliography]{revtex4-1}
\usepackage{url,ulem}
\usepackage{amsmath,amssymb}
\usepackage[dvipdfmx]{graphicx}
\usepackage{color,verbatim}

\newcommand{\dfracp}[2]{\dfrac{\partial #1}{\partial #2}}

\begin{document}
\title{Towards a classification of bifurcations in Vlasov equations}
\author{J. Barr\'e}
\affiliation{Institut Denis Poisson, Universit\'e d'Orl\'eans, CNRS, Universit\'e de Tours, France;
and Institut Universitaire de France}
\author{D. M\'etivier}
\affiliation{Center for Nonlinear Studies and Theoretical Division T-4 of Los Alamos National Laboratory, NM 87544, USA}
\author {Y. Y. Yamaguchi}
\affiliation{Department of Applied Mathematics and Physics, Graduate School of Informatics, Kyoto University, Kyoto 606-8501, Japan}

\begin{abstract}
We propose a classification of bifurcations of Vlasov equations, based on the strength of the resonance between the unstable mode and the continuous spectrum on the imaginary axis. We then identify and characterize a new type of generic bifurcation where this resonance is weak, but the unstable mode couples with the Casimirs, which are constants of motion, to form a size 3 Jordan block. We derive a three-dimensional reduced noncanonical Hamiltonian system describing this bifurcation: coupling with the Casimirs controls the phase space portrait. Comparison of the reduced dynamics with direct numerical simulations on a test case gives excellent agreement. We finally discuss the relevance of this bifurcation to specific physical situations. 
\end{abstract}

\maketitle

A wide variety of physical systems are governed over certain time scales by mean-field forces rather than collisions
between their constituants. The appropriate kinetic description is then a Vlasov, or Vlasov-like, equation.
These equations possess both regular features (such as an infinite number of conserved quantities) and chaotic 
ones (such as the development of infinitely fine structures in phase space) which make both the understanding of their 
qualitative behavior and their numerical simulation famously difficult problems, relevant in various physical fields.
In particular, Vlasov-like equations have an uncountable number of stationary states, and linear and nonlinear stability studies of these 
states have led, among other important physical and mathematical concepts, to the discovery of Landau damping \cite{Landau46} close to stable stationary states.
We are concerned in this article with the question: What happens close to \emph{weakly unstable} stationary states? This amounts to a study of local 
bifurcations of Vlasov equations. The rationale is that these bifurcations i) should be universal, i.e. be relevant for all types of Vlasov 
equations, and ii) could provide basic building blocks to describe the qualitative behavior of these equations. 

Vlasov equations are Hamiltonian systems,
and while bifurcations in Hamiltonian systems are well known and classified
\cite{Meyer74,Meyer75,golubitsky-stewart-87},
the specificities of Vlasov equations bring difficulties.
First,
their Hamiltonian structure is noncanonical, and highly degenerate \cite{MorrisonGreene80,Morrison80,Morrison98}, which is the origin of the infinite number of conserved quantities, called Casimir invariants.
Second, the linearized Vlasov evolution typically features continuous spectrum on the imaginary axis, and a 
growing unstable mode can create resonances with part of this spectrum, triggering complex dynamics.

The study of Vlasov bifurcations is an old topic.
One of the most common Vlasov bifurcations describes the destabilization of a homogeneous stationary state, and is relevant in plasma physics (bump on tail, 
or two beams instabilities) and fluid dynamics (shear flow instability);
the nonlinear development of the instability gave rise to a debate 
starting in the 60s \cite{Dawson61,Frieman63,Simon76,Denavit85}
and concluded in the 90s, when it became clear that it was governed by resonances. This instability is characterized by 
strong nonlinear effects, at the origin of "trapping scaling" \cite{Dawson61,Crawford94,Crawford95}: the amplitude of the unstable mode saturates at a level $O(\lambda^{2})$, where $\lambda$ is the instability rate, rather than the much larger $O(\lambda^{1/2})$, as happens for standard supercritical pitchfork or Hopf bifurcations. Furthermore, an infinite dimensional "normal form" has been derived for the near 
threshold dynamics, called the Single Wave Model (SWM) \cite{ONeil71,delcastillonegrete-98,balmforth-morrison-thiffeault-13}.
It is clear that however important this SWM example may be, it is just one case among many other possible bifurcations, which are much less studied. 

As a first example beyond the SWM case, a kind of "very strong" resonance has already been identified in the literature \cite{crawford-jayaraman-96,crawford-jayaraman-99,Balmforth}.
The critical eigenvector associated with the instability
  is in this case singular and entails stronger nonlinear effects altering the trapping scaling characteristic of the SWM bifurcation: nonlinear 
  effects kick in at an amplitude $O(\lambda^{5/2})$, to be compared with the $O(\lambda^2)$ for trapping scaling. While this bifurcation was only studied for two species plasmas, it can likely be found in other contexts as well. 

As a second example, some instabilities do not give rise to resonances, or only to weak ones. The weak resonance
happens for homogeneous stationary states with some 
special velocity distributions~\cite{balmforth-morrison-thiffeault-13},
and, more importantly, this is a generic situation for nonhomogeneous stationary states with real bifurcating eigenvalue, as shown in~\cite{BMY}. 
At first sight, it seems that such nonresonant bifurcations can be studied through standard 
center manifold reduction and hence would fall into
the class of normal finite dimensional canonical Hamiltonian bifurcations. While it is true in some cases \cite{balmforth-morrison-thiffeault-13},
we highlight in this article a new type of generic bifurcation for weakly resonant non-oscillatory instabilities:
a neutral mode associated with a Casimir invariant combines with a stable and an unstable modes, thereby forms a three-dimensional Jordan block, and controls the bifurcation.
In particular, when the initial perturbation does not modify the Casimirs, nonlinear effects kick in when the unstable mode reaches an amplitude $O(\lambda^2)$. However, 
this scaling can be strongly modified by the Casimir coupling, see Table~\ref{tab:classification}.
In \cite{BMY}, this bifurcation is studied using an unstable manifold expansion "\`a la Crawford" \cite{Crawford94}, which assumes a perturbation exactly along the unstable direction, i.e. $Z=0$. The 3D reduced model derived below describes quantitatively the dynamics for any initial perturbation.

\begin{table}[t]
  \centering
  \caption{Classification sketch for bifurcations in Vlasov-like systems.
    VSR (SR,WR) represents  (very) strong (weak) resonance
    which occurs without (no-C) or with (w-C) coupling
    with the Casimir modes at the linear level.
    SWM represents the single wave model.
    SE (RE) represents singular (regular) eigenvectors.
    The "Scaling" column precises the perturbation amplitude at which 
    non linear effects kick in; $\lambda$ is the instability rate. $Z$ is the Casimir perturbation value.
    The fourth line is highlighted, as the main subject of this work.  
  }

  \begin{tabular}{llllll}
    \hline
    Resonance & Casimirs & Reduction & E.Vec. & Scaling & Ref. \\
    \hline
    VSR & no-C & See \cite{Balmforth} & SE & $\lambda^{5/2}$ & \cite{crawford-jayaraman-96,crawford-jayaraman-99,Balmforth}\\
    SR & no-C & SWM & RE & $\lambda^2$ & \cite{ONeil71,delcastillonegrete-98,balmforth-morrison-thiffeault-13}  \\
    WR & no-C &  Finite dim. & RE &$\lambda^{1/2}$ & \hspace{-0.9cm}\cite{balmforth-morrison-thiffeault-13} Sec. IVA \\
    {\bf WR} & {\bf w-C} & {\bf Finite dim.} & {\bf RE} & $\mathbf{(\lambda^4+Z)^{1/2}}$ & $\substack{ {\bf This}\\{\bf work}}$ \\
    \hline
  \end{tabular}
  \label{tab:classification}
\end{table}
Summarizing the above discussion,
  Table \ref{tab:classification} sketches a classification of bifurcations
  in Vlasov-like equations.
  It is the first product of this article.
In the following, we turn to the main contribution of this work: we will first show that the scenario involving the weak resonance and coupling with the Casimirs is generic, and study it at linear and nonlinear levels, until we obtain a reduced three-dimensional Hamiltonian which plays the role of a normal form for this bifurcation. We then provide an illustration in a spatially one-dimensional model,
where all computations can be performed explicitly and the 
reduced dynamics can be quantitatively compared with direct numerical simulations of the Vlasov equation.

\emph{Dimension reduction: linear analysis-}
Vlasov equation for the phase space density $F(q,p,t)$ describes the evolution of the phase space density, and reads
\begin{equation}
  \dfracp{F}{t} + \dfracp{H[F]}{p} \dfracp{F}{q}
  - \dfracp{H[F]}{q} \dfracp{F}{p} = 0,
  \label{eq:Vlasov}
\end{equation}
where $(q,p)$ are phase space variables, $H[F]=p^{2}/2+V[F] + \text{Cst}$ is the one-body Hamiltonian with
\begin{equation}
  V[F](q,t)=\int v(q-q') F(q',p',t)dq'dp'
\end{equation}
and $v(q)$ is the two-body potential. Casimirs are conserved functionals of the type $\int \varphi[F(q,p,t)]dqdp$, for any smooth function $\varphi$. It is well known that \eqref{eq:Vlasov} can be seen as an infinite dimensional non canonical Hamiltonian system, and that Casimir conservation directly stems from the degeneracy of this structure \cite{MorrisonGreene80,Morrison80}.
The three-dimensional Jordan block structure at the bifurcation point is not related to the infinite dimensionality. Hence, to make our point clearer, we  
will use a matrix formal representation of Vlasov equation as a noncanonical Hamiltonian system:
\begin{equation}
  \label{eq:finite-Poisson}
  \dot{y}
  = J(y)\nabla H(y) 
\end{equation}
where $J$ is a degenerate Poisson matrix,
depending on the state $y$, and $H$ is the Hamiltonian. 
For the Vlasov equation, $y$ would be the density function $F$ over phase space, the gradient
$\nabla$ a functional derivative, and $J$ an operator.
Our setting is as follows: we consider a family of stationary state
$\{y_{\mu}\}$, where $\mu=0$ is the critical point at which the stability changes,
and $\{y_{\mu}\}$ are close to $y_{0}$ for $\mu$ small.
We reduce Eq.~\eqref{eq:finite-Poisson} by projecting it
on a lower dimensional space governing the slow dynamics,
which is extracted from the linearized equation.
We assume that the imaginary part of the unstable eigenvalue is zero
(steady state bifurcation),
otherwise a coupling with Casimir invariants,
which at the linear level correspond to zero modes, is impossible.
We also assume no resonance, or weak resonance, with the continuous spectrum 
on the imaginary axis: hence one expects that a reduction to an effective 
finite dimensional dynamics is possible close to $\mu=0$. 

We assume that $y_{0}$, the stationary point of interest,
is not singular for $J$,
i.e. $J$ has a constant rank in $V(y_{0})$,
a neighborhood of $y_{0}$.
Weinstein's splitting theorem \cite{weinstein-83} implies that, up to a local coordinate change, the Poisson operator can be written as
\begin{equation}
  J =
  \begin{pmatrix}
    J_{0} & O_{2n,m} \\
    O_{m,2n} & O_{m,m} \\
  \end{pmatrix},
  \quad
  J_{0} =
  \begin{pmatrix}
    O_{n,n} & I_{n} \\
    -I_{n} & O_{n,n} \\
  \end{pmatrix}
\end{equation}
in $V(y_{0})$ including a part of the family of stationary states $\{y_{\mu}\}$.
Here, $O_{k,l}$ is the zero matrix of size $k\times l$
and $I_{n}$ is the unit matrix of size $n$.
The appropriate change of variable can be built order by order, see
\cite{MorrisonVanneste}, where the procedure is called "beatification". In practice, we will only 
need the lowest order.
The degenerate part, $O_{m,m}$ in $J$, corresponds to the Casimir invariants $z$
with the notation $y=(x,z)~(x\in\mathbb{R}^{2n},z\in\mathbb{R}^{m})$.
This part makes $J$ not invertible: in particular we may have $\nabla H(y_{0})\neq 0$ even at the stationary point $y_{0}$.
However we know $\nabla H(y_{0})$ belongs to the kernel of $J$, which is spanned by the linearized Casimir invariants; hence we may assume
$\nabla H(y_{0})=0$ by adding to $H$
a linear combination of the Casimir invariants.
The linearized equation around $y_{\mu}$ is, therefore,
$\dot{\eta}_{\mu}=L_{\mu}\eta_{\mu}$,
where $\eta_{\mu}=y-y_{\mu}$, $L_{\mu}=JS_{\mu}$, and $S_{\mu}$ is the Hessian matrix of $H$ at $y_{\mu}$.

The linearized dynamics at the critical stationary state $y_0$ is given by the linear operator $L_0$.
By assumption, $L_0$ has $0$ as eigenvalue, and, using the adjoint $L_0^\dagger$, we want to build the projection onto the associated 
eigenspace, which may contain generalized eigenvectors.
Indeed our first result is to show that $L_0$ generically has 
a three-dimensional Jordan block associated with the generalized eigenvalue $0$: if $\psi_0$
is an eigenvector, i.e. $L_0\psi_0 =0$, there exist $\psi_1$ and $\psi_2$ such that
\[
L_0\psi_1=\psi_0~,~L_0\psi_2=\psi_1.
\] 
The following proof by the matrix formalism is justified
by the weak resonance condition
(see Supplemental Material (SM) \cite{SM}
for an example in an explicitly infinite setting).
Omitting the subscript $0$ of $S_0$ and $L_{0}$,
we write the linearized equation at $y_{0}$ as
\begin{equation}
  \dot{\psi} = JS\psi =: L\psi, \quad
  S =
  \begin{pmatrix}
    S_{xx} & S_{xz} \\
    S_{zx} & S_{zz} \\
  \end{pmatrix},
\end{equation}
where $S_{xx}\in\mathbb{R}^{2n\times 2n}$ and $S_{zz}\in\mathbb{R}^{m\times m}$.
Clearly, ${\rm rank}(L)\leq 2n$.
Furthermore, ${\rm rank}(S_{xx})<2n$
because $y_{0}$ is a critical stationary point at which the stability changes.
The generic case gives ${\rm rank}(S_{xx})=2n-1$
together with ${\rm rank}(L)=2n$.
Denoting the inner product on $\mathbb{R}^{2n}$ (resp. $\mathbb{R}^{2n+m}$)
by $\langle\cdot,\cdot\rangle_n$
(resp. $\langle\cdot,\cdot\rangle_{n,m}$)
  and $\psi=(\xi,\zeta)~(\xi\in\mathbb{R}^{2n},\zeta\in\mathbb{R}^{m})$,
we make two remarks:
  i) The equation $L\psi=v$ has a solution if and only if
  $\langle \varphi,v\rangle_{n,m}=0$ for any
  $\varphi \in{\rm Ker} L^\dagger$,
  where $L^{\dagger}$ is the adjoint operator (transposition) of $L$.
  Since ${\rm Ker} L^\dagger=\{0_{2n}\}\times \mathbb{R}^m$,
  where $0_{2n}$ is the origin of $\mathbb{R}^{2n}$,
  the above equation has a solution if and only if
  the last $m$ coordinates of $v$ vanish.
  ii) Noting $(J_{0}S_{xx})^{\dagger}=-S_{xx}J_{0}$
  and $\dim{\rm Ker}(S_{xx})=1$,
the equation $J_0S_{xx} \xi =w$ has
a solution if and only if $\langle J_0 \xi_0,w\rangle_n=0$,
where $\xi_{0}$ is a vector spanning ${\rm Ker}( S_{xx})$.

The critical eigenvector is $\psi_0=(\xi_0,0_m)$.
By i) the equation $L\psi_1=\psi_0$ has a solution. 
This equation writes for $\psi_1=(\xi_1,\zeta_1)$:
\begin{equation}
  J_0S_{xx} \xi_1 +J_0S_{xz} \zeta_1 = \xi_0.  
\end{equation}
Since $\langle J_0 \xi_0,\xi_0\rangle_n=0$,
by ii) the above equation has a solution with $\zeta_1=0_m$,
and we have found the first generalized eigenvector $\psi_1=(\xi_1,0_m)$.
Again by i) $L\psi_2=\psi_1$ has a solution
because the last $m$ coordinates of $\psi_1$ vanish. However, in general $\langle J_0 \xi_0,\xi_1\rangle_n \neq 0$, hence
by ii) it is 
impossible to find $\psi_2$ with a vanishing second component:
the solution is of the form $\psi_2=(\xi_2,\zeta_2\neq 0_m)$.
It is now clear, by i) again, that $L\psi=\psi_2$ has no solution, and there are only two generalized eigenvectors, forming a size $3$ Jordan block.
Furthermore, $\psi_2$ has a non zero component $\zeta_{2}$
along the direction of the Casimirs,
this is the specificity of this bifurcation.
If the Hamiltonian does not induce any coupling with the Casimir modes at the linear level, i.e. $S_{xz}=O_{2n,m}$,
the assumption ${\rm rank}(L)=2n$ breaks down, and ${\rm rank}(L)=2n-1$ instead. Generically $L$ then has a size 2 Jordan block, without coupling with the Casimir modes,
because ${\rm Ker}L^{\dag}$ has one more dimension.

\emph{Dimension reduction: non linear analysis-}
We now study
the bifurcation at the nonlinear level
by projecting the infinite dimensional dynamics onto
$E_{0}={\rm Span}(\psi_{0},\psi_{1},\psi_{2})$.
We build an invariant three-dimensional 
manifold whose tangent space at $y=y_{0}$ is $E_{0}$,
through a local Taylor expansion
together with the expansion on the bifurcation parameter $\mu$. 
A point (i.e. a phase space function) close to $y_0$ on this manifold can then be written as
\begin{equation}
  \label{eq:manifold}
  y=y_0 + \sum_{i=0}^{2} A_i \psi_i + 
  O_{2}(A_{0},A_{1},A_{2}),
\end{equation}
where the order $2$ remainder term, $O_{2}(A_{0},A_{1},A_{2})$,
describes the manifold's "local curvature". Our goal is to describe the dynamics on this manifold, 
that is the time evolution of the $A_i$'s. Notice that the nonlinearity of the Vlasov equation is quadratic, hence when applied to 
$y-y_0$ as in \eqref{eq:manifold}, the curvature term only produces terms of order $A^3$ or higher. Thus, at leading nonlinear order 
$A^2$, it is enough to approximate the manifold by $E_0$.
We then take an initial condition $y(t=0)=y_0+\sum_{i=0}^{2} A_i(0) \psi_i$; up to quadratic order in the $A_i$'s, the evolution is 
$y(t)=y_0+\sum_{i=0}^{2} A_i(t) \psi_i$, and we 
aim to determine the $A_i(t)$.
The strategy is: 
i) restrict the Poisson structure to the subspace $E_0$;
ii) expand and truncate the Hamiltonian up to cubic order in the $A_{i}$s.
After appropriate changes of variables 
$(A_{i})\to u=(u_{0},u_{1},u_{2})$,
we are left with the reduced dynamics
$\dot{u}=J_{\rm red}\nabla H(u,\mu)$ with
\begin{equation}
\label{eq:Jred}
J_{\rm red}=\begin{pmatrix}
    0 & 1 & 0 \\
    -1& 0 & 0 \\
    0 & 0 & 0 \\
  \end{pmatrix}~,~
  H(u,\mu) = H_{2}(u,\mu) + H_{3}(u),
\end{equation}
where $H_{k}$ are homogeneous polynomials of degree $k$ in its $u$ variables.
We keep only the leading terms in $\mu$, of order $\mu u^{2}$.

\begin{figure}[htbp]
\includegraphics[width=0.48\textwidth]{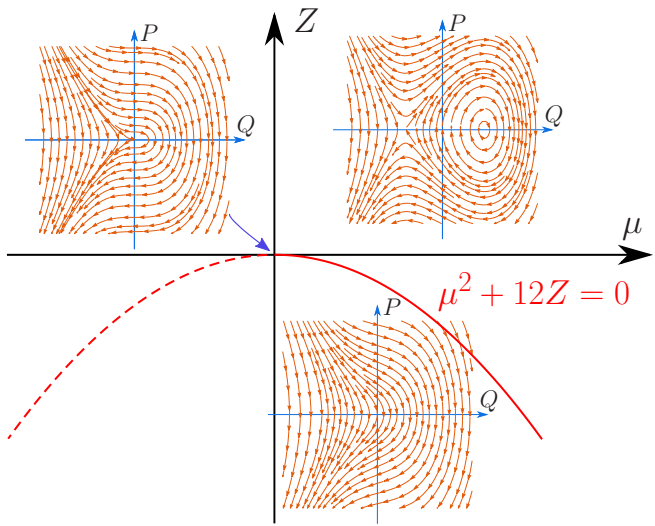}
\caption{\label{fig:main}
  Sketch of the phase space portraits according to the reduced Hamiltonian. Since $Z$ is a conserved quantity depending on the initial perturbation, it can be thought of as another parameter controlling the bifurcation. For $\mu^{2}+12Z>0$, $\partial\Phi/\partial Q=0$
has two real solutions corresponding to one stable and one unstable stationary states. Depending on the initial condition, trajectories in the $(Q,P)$ plane can be trapped around the stable state, or unbounded, eventually leaving the perturbative regime. For $\mu^{2}+12Z<0$, there is no stationary state, and all trajectories are unbounded. At variance with the finite dimensional cases, when $(Q,P,Z)=(0,0,0)$ is a stable stationary state (i.e. $\mu<0$), there are not necessarily purely imaginary eigenvalues close to $0$, and the reduced Hamiltonian may not be meaningful. 
    }
\end{figure}

\emph{Normal form of the bifurcation-}
Our last task is to provide a normal form for the reduced Hamiltonian, i.e.
make it as simple as possible through changes of variables.
To recover the size $3$ Jordan block at $\mu=0$ from
$J_{\rm red}\nabla H_{2}$,
we set the quadratic Hamiltonian $H_{2}$ as
\begin{equation}
  \label{eq:Hred-quad}
  H_{2}(u,\mu) = (u_{1}^{2} - \mu u_{0}^{2})/2 - u_{0} u_{2}.
\end{equation}
The parameter $\mu$ controls the stability of the origin.
The idea to derive the normal form is to transform the coordinates
to $U=(Q,P,Z)$, defined as $u=U+T(U)$,
in order to simplify $H(U+T(U))=:\bar{H}(U)$.
We assume that $T(U)$ is a homogeneous polynomial of order 2
and thus has $6\times 3=18$ parameters.
Imposing to keep the standard form 
\eqref{eq:Jred} for $J_{\rm red}$ reduces the number of free parameters to $10$.
The cubic term of $\bar{H}(U)$ is $H_{3}(U)+T\cdot\nabla H_{2}(U)$.
All terms in $H_{3}(U)$ can be eliminated by appropriate choices
of the $10$ free parameters left in $T$, 
except the $Q^{3}$ and the $ZQ^2$ ones, see SM \cite{SM}.
Moreover, the coefficient of the remaining $Q^{3}$ term can be scaled to $1$, and, 
since $Z$ is conserved by the dynamics, the $ZQ^2$ term can be absorbed in a redefinition of $\mu$.
Consequently, around the critical point,
the normal form of the reduced Hamiltonian is
\begin{equation}
  H_{\rm red} = P^{2}/2 + \Phi(Q,Z), 
  \quad
  \Phi = -\mu Q^2/2- QZ + Q^{3}
\label{eq:ham_red}
\end{equation}
up to cubic order.
This Hamiltonian provides a kind of three-dimensional
``fish''-shape bifurcation
\cite{golubitsky-stewart-87},
with an important observation: the value of the Casimir invariant,
$Z$, controls the bifurcation; the details are shown on Fig.~\ref{fig:main}.
Note that if $Z=0$ the stable fixed point appears at
distance of order $Q^{\rm sfp}\propto \lambda^2 \sim \mu$ from the reference stationary state, but even a small $Z$ (of order $\lambda^4$) modifies this scaling $Q^{\rm{sfp}}\propto \sqrt{Z+\mu^2}$, see Eq.(53) in \cite{SM} and following comments.

We conclude that under the hypotheses:
i) steady state bifurcation and ii) weak resonance with the continuous spectrum,
Fig.~\ref{fig:main} describes a new type of generic bifurcation expected in Vlasov-like systems. We turn now to explicit 
computations in Vlasov equations in a two-dimensional phase space to demonstrate that this bifurcation indeed occurs and accurately describes 
the dynamical behavior of the system in the vicinity of the bifurcation point.

\emph{Explicit example-}
We consider a periodic domain $\mathbb{T} =[0,2\pi[$ in the space variable $q$,
hence the phase space is $(q,p)\in \mathbb{T}\times \mathbb{R}$.
The Vlasov equation reads \eqref{eq:Vlasov} for the density $F(q,p,t)$.
To simplify explicit computations and numerical simulations,
we will use the two-body potential $v(q)=-\cos q$, i.e. the so-called
Hamiltonian mean-field (HMF)
model \cite{LyndenBell,inagaki-konishi-93,Antoni}.

We take the family of ``Fermi-Dirac'' stationary states
\begin{equation}
  \label{eq:Fermi-Dirac}
  F_\mu(J)= \mathcal{N}^{-1}\frac{1}{1+e^{\beta[H(J)-(\mu-\kappa)]}},
\end{equation}
where $\mathcal{N}^{-1}$ is the normalization factor,
$\mu$ controls the bifurcation,
and $\kappa(\beta)$ is chosen so that the critical point is $\mu=0$.
The rotational symmetry of the HMF model permits to write
the stationary potential as $V[F_{\mu}]=-M_{\mu}\cos q$
without loss of generality,
where $M_{\mu}=M[F_{\mu}]=\int \cos q F_{\mu}(J) dqdp \neq 0$
is the stationary magnetization.
The action variable $J$ is defined for the pendulum Hamiltonian
$H_{\mu}=H[F_{\mu}]=p^{2}/2+M_{\mu}(1-\cos q)$.
The family \eqref{eq:Fermi-Dirac} undergoes a bifurcation \cite{BMY}:
$F_{\mu <0}$ is stable, and $F_{\mu>0}$ is unstable.
The trajectories for $H_{\mu}$ are oscillating around
$(q,p)=(0,0)$ or rotating along the torus $\mathbb{T}$,
their frequency is $\Omega_{\mu}(J)=dH_{\mu}/dJ$.
The definitions of $J$ and $\Omega_{\mu}$ depend on $\mu$ through $M_{\mu}$;
however, this dependence does not enter in the equations up to the second
order, hence we may use $\mu=0$ for the definitions of these quantities.
As pointed out in \cite{BMY}, trajectories with near zero frequency
correspond to trajectories close  to the separatrix; since the frequency vanishes only
logarithmically close to $J=J_c$, the action at the separatrix, their density is small: in particular, 
$1/\Omega_0(J)$ is integrable near $J=J_c$.
This small density ensures that the weak resonance condition is satisfied, see \cite{SM}.

\begin{figure}
\includegraphics[width=0.48\textwidth]{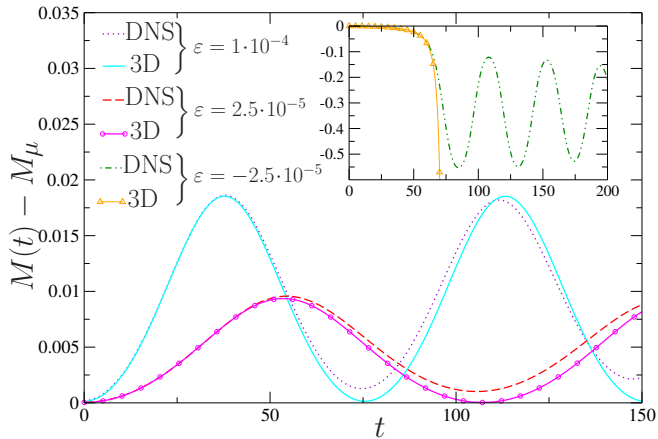}
\caption{\label{fig:simulations_hmf}
  Temporal evolution of magnetization:
  comparison between DNS of 
  \eqref{eq:Vlasov} (dashed and dotted lines)
  and the analytic solution (3D) of Fig.~\ref{fig:main} (solid lines).
Two different amplitudes for the initial perturbation are used.
In the inset we show the DNS and the reduced dynamics for the same initial state with $\varepsilon<0$ (resulting in $\mu^2+12Z<0$);
 the 3D dynamics diverges, while the DNS shows large scale oscillations. Details of the numerical simulations can be found in the main text and SM \cite{SM}.
 }
\end{figure}

As described in \cite{SM},
we start from the Vlasov dynamics 
\eqref{eq:Vlasov}
in a neighborhood of the critical stationary state $F_0$,
and reduce it to Hamiltonian \eqref{eq:ham_red},
with explicit expressions of the changes of variables, coefficients and initial conditions involved: hence, without any adjustable parameter, 
we can directly and quantitatively compare the predictions of Fig.~\ref{fig:main} with numerical simulations of 
\eqref{eq:Vlasov}.
This comparison is presented on Fig.~\ref{fig:simulations_hmf}. 
We have used the initial condition
\begin{equation} 
F(t=0) = F_\mu +\varepsilon \cos q e^{-\beta_T p^2},  
\end{equation}
with a parameter $\varepsilon$ controlling the sign and amplitude of the initial perturbation.
This initial perturbation has zero projection on $\psi_1$, but has contributions of order $\varepsilon$ along both $\psi_0$ (the unstable direction) and $\psi_2$ (the Casimir perturbation direction). The latter contribution gives $Z=O(\varepsilon)$, while the former puts the initial point at distance $O(Z)$ from the origin in the $(Q,P)$ plane. Since the stable fixed point is at distance $O(\sqrt{Z+\lambda^4})$ from the origin, the dynamics on the $(Q,P)$ plane strongly depends on $Z$. 

The Vlasov simulations are performed using the algorithm of \cite{Rocha}, and we use the analytic solution of the reduced dynamics in terms of Weierstrass $\wp-$function~\cite{SM}.
We give four remarks. 
i) The agreement is good, both in terms of frequency and amplitude,
over fairly long time scales.
ii) There is a small damping (and frequency shift)
acting on the direct numerical simulation (DNS),
an effect that we attribute to the numerical dissipation well known
in Vlasov simulations \cite{banks2010new,silantyev2017langmuir}. We confirmed that hypothesis by varying grid sizes (see SM \cite{SM}). However, one cannot exclude the possibility of a weak Landau damping-like effect not described by the reduced Hamiltonian.
iii) Changing the initial perturbation amplitude has an important 
effect on the dynamics: this is a signature of the importance
of the $Z$ coordinate, representing the coupling with the Casimirs.
iv) The Vlasov dynamics indeed leaves the perturbative regime when predicted by the reduced model (see inset). The observed large scale oscillations suggest the existence of a periodic solution; they are out of reach of the reduced model, but it is worth mentioning that a reduced Hamiltonian at fourth order indeed predicts in some cases the confinement of trajectories and large scale oscillations \cite{SM}.

\emph{Physical examples and discussion-}
To summarize, we have first identified and described on general theoretical grounds a new type of bifurcation for Vlasov systems, and then proved that it indeed occurs on a simple system. Including the coupling between the instability and the Casimirs proved critical to accurately describe the bifurcation. We discuss now its possible relevance in more realistic physical systems:
we need to find situations where the basic conditions of the weak resonance and possible coupling with Casimir modes are satisfied.  
\begin{itemize}

\item Instabilities of Bernstein-Greene-Kruskal 
modes in plasmas provide a vast class of natural candidates. The simplest cases, based on the 1D Vlasov-Poisson equation, are similar to the HMF example studied above, and we know that bifurcations do occur (see for instance \cite{PankavichAllen2014}): we expect some of these bifurcations to be described by the theory put forward in this paper.  

\item 
Radial orbit instability is well known in astrophysics (see for instance \cite{BinneyTremaine}), and believed to play a role in determining the structure of galaxies. It occurs in self-gravitating systems, when the amount of particles (usually stars) with small angular momentum increases.  The nonlinear analysis in \cite{Palmer} suggests similarities  with the phenomenology of Fig.~\ref{fig:main}: in particular, the instability is non-oscillating, and, depending on the initial perturbation, the saturated state may be close to the reference stationary state, or far away. Still in astrophysics, gravitational loss cone instability (see for instance \cite{polyachenko-polyachenko-shukhman-07}) could also present a similar phenomenology, however we are not aware of a nonlinear analysis of this situation. 

\item  A class of Hamiltonian models of oscillators synchronization is introduced in \cite{Smereka98}: identical nonlinear oscillators are coupled through a mean-field, and the 
stationary state with all oscillators desynchronized can undergo a bifurcation towards a synchronized state. We believe this bifurcation should be in some cases similar to the one described in this article, with the caveat that no coupling with Casimirs takes place; accordingly, the Jordan block at criticality has only size two. This type of models can for instance describe coupled electric circuits \cite{Smereka98}, as well as, somewhat unexpectedly, pressure waves in bubbly fluids \cite{Smereka96,smereka-02}.  

\end{itemize}

We conclude that many physical systems from very different fields can be expected to follow the phenomenology in Fig.~\ref{fig:main}; specific studies and simulations in each case are now needed to confirm or infirm these predictions, and assess their physical importance. 

\begin{acknowledgments}
D.M. carried the work at LANL under the auspices of the National Nuclear Security Administration of the U.S. Department of Energy under Contract No. DE-AC52-06NA25396 and was partially supported by LANL/LDRD/CNLS projects. 
Y.Y.Y. acknowledges the support of JSPS KAKENHI Grant No. 16K05472.
\end{acknowledgments}

\end{document}